\documentclass[11pt]{article}

\usepackage{amsmath}
\usepackage{xcolor}
\usepackage{natbib}
\usepackage{graphicx}
\usepackage{hyperref}

\usepackage{fullpage}

\usepackage{booktabs} 
\usepackage[ruled]{algorithm2e} 

\SetAlFnt{\small}
\SetAlCapFnt{\small}
\SetAlCapNameFnt{\small}
\SetAlCapHSkip{0pt}
\IncMargin{-\parindent}

\usepackage[normalem]{ulem}
\usepackage{cleveref}
\definecolor{mylightblue}{HTML}{6497b1}
\definecolor{mynavyblue}{HTML}{011f4b}
\definecolor{lightblue}{HTML}{5067FD}

\DeclareMathOperator*{\argmax}{arg\,max}

\title{Behavioral Study of Dashboard Mechanisms}

\author{Paula Kayongo \footnote{Department of Computer Science, Northwestern University.
Email: \texttt{paulakayongo2023@u.northwestern.edu}}
\and Jessica Hullman\footnote{Department of Computer Science, Northwestern University.
Email: \texttt{jhullman@northwestern.edu}}
\and Jason Hartline\footnote{Department of Computer Science, Northwestern University.
Email: \texttt{hartline@northwestern.edu}}
}

\begin{document}


\maketitle

\begin{abstract}

Visualization dashboards are increasingly used in strategic settings like auctions to enhance decision-making and reduce strategic confusion. This paper presents behavioral experiments evaluating how different dashboard designs affect bid optimization in reverse first-price auctions. Additionally, we assess how dashboard designs impact the auction designer's ability to accurately infer bidders' preferences within the dashboard mechanism framework. We compare visualizations of the bid allocation rule—commonly deployed in practice—to alternatives that display expected utility. We find that utility-based visualizations significantly improve bidding by reducing cognitive demands on bidders. However, even with improved dashboards, bidders systematically under-shade their bids, driven by an implicit preference for certain wins in uncertain settings. As a result, dashboard-based mechanisms that assume fully rational or risk-neutral bidder responses to dashboards can produce significant estimation errors when inferring private preferences, which may lead to suboptimal allocations in practice. Explicitly modeling agents' behavioral responses to dashboards substantially improves inference accuracy, highlighting the need to align visualization design and econometric inference assumptions in practice.

\end{abstract}

\vspace{1cm}
\setcounter{tocdepth}{1} 



\section{Introducton}

A key challenge to obtaining good outcomes in strategic decision-making settings, e.g., online auctions \citep{cox1983test}, labor markets \citep{afeche2022ride}, and school matching markets \citep{rees2018suboptimal,pathak2008leveling, dur2018identifying} arises from the fact that decision-makers~({\em henceforth agents}) often struggle to identify effective decision-making strategies. To encourage more efficient strategic behavior that leads to improved resource allocation, market designers~({\em henceforth principals}) commonly deploy {\em visualization dashboards} that communicate payoff-relevant market information to the agents. For instance, the Google AdWords auction dashboard visualizes a line chart that depicts predicted click volume and cost per click as a function of advertiser bids. This paper conducts behavioral studies to experimentally evaluate different dashboard designs and their effiacy at inducing more optimal bidding in first-price single-item winner-pays-bid auctions.

Systematic deviations from optimal bidding (i.e., the exact right amount of bid shading) in first-price auctions have been extensively documented in the literature.  
A classic finding by \citet{cox1983test} is that bidders in first-price auctions often undershade, submitting bids that exceed those expected under risk-neutral Bayes-Nash equilibrium. 
Undershading in first-price auctions has been attributed to unusual mental models \cite{eyster2005cursed,crawford2007level,niederle2023cognitive}, computational constraints~\citep{dorsey2003explaining, levin2016separating}, and unusual preferences~\citep{cox1985experimental}.  In first-price auctions, the bidding process involves several steps, including understanding the competitors' value distribution, deducing how it leads to a specific bidding pattern, and how this pattern determines the allocation rule, and finally, optimizing one's bid based on this rule. Compared to this extensive behavioral literature on bidding in auctions, our study removes the problem of reasoning about how other bidders’ strategize and asks whether bidders can respond optimally given fixed competition and what format of the visual depictions of competition should take to best aid in good bidding.

A recent line of work~\citep{hartline2019dashboard} demonstrates that good mechanisms can be designed around the idea that bidders respond optimally to dashboards.
Specifically, the {\em dashboard mechanism } of \citet{hartline2019dashboard} and \citet{deng2021welfare}, incentivize bidders to rationally optimize according to the dashboard and efficiently allocate resources when they do.
The dashboard mechanism approach can be viewed as a reduction from a many-agent allocation problem to the problem of each agent bidding against a single opponent with a known bid distribution, which is visualized in the dashboard. 
A critical step in dashboard mechanisms is the use of econometric inference to infer the bidders’ private values given their bids.
\citet{hartline2019dashboard} and~\citet{deng2021welfare} assume rational agents that perfectly optimize their
bids against the dashboard when conducting this econometric inference, and therefore, the bidders’
values can be exactly inferred. 
One goal of this paper is to understand how behavioral bidders deviate from the rational bidding assumption and the extent to which these deviations affect bid inference in dashboard mechanism design.

\begin{figure*}
  \includegraphics[width=\textwidth]{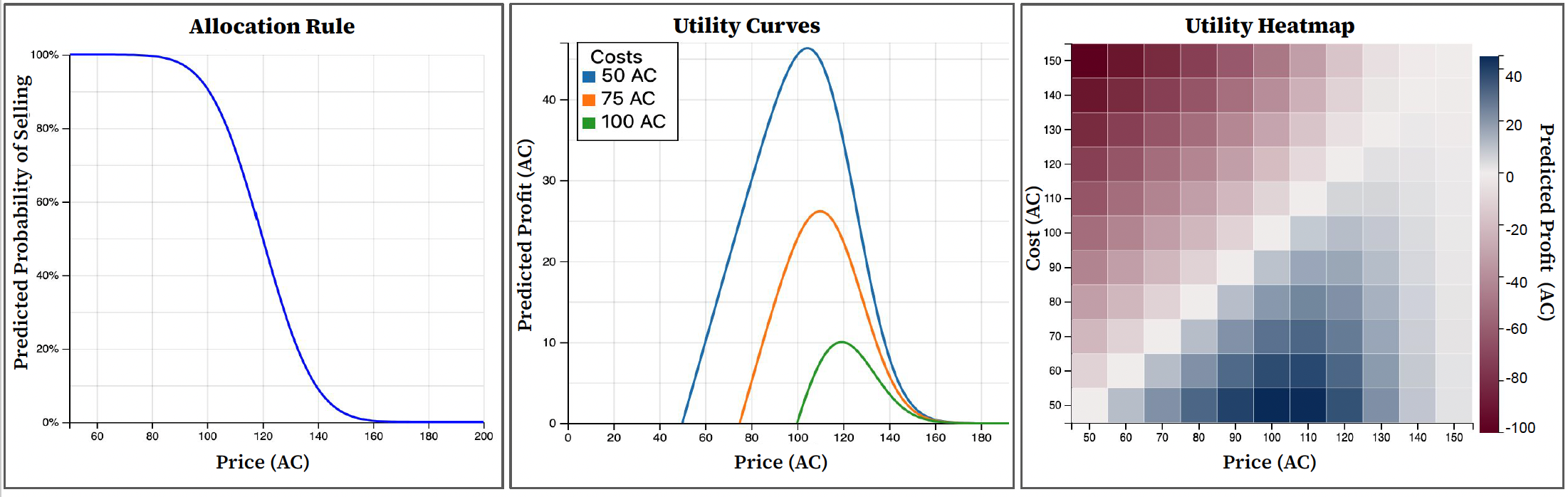}
  \caption{Visualization designs evaluated in behavioral experiments}
  \label{fig:teaser}
\end{figure*}

To evaluate how dashboard design influences bidding behavior and the practical feasibility of dashboard mechanisms, we contribute the results of a large-scale reverse first-price experimental auction. The reverse auction differs from the forward auction in that, while in forward auctions, buyers with private values bid to acquire items, in reverse auctions, sellers with private costs bid to provide goods.\footnote{In auction experiments like the one we conduct, to establish normative behavior, participants need to be endowed with values or costs. We endow costs (of sellers in a reverse auction) because it is more natural to convey to study participants.} In each auction round, we provide
participants with payoff-relevant visualizations, solicit bids, determine outcomes, and provide
feedback.
We vary three visualization types between subjects, chosen to vary in the extent to which they support direct reasoning about bids conditional on endowed costs (see \Cref{fig:teaser}): 
\begin{description}
    \item[Allocation Rule] depicts the probability of selling an item as a function of the seller's bid. This visualization is similar to the one deployed in practice in dashboards such as Google Adwords.
    \item[Utility Curves] depicts the expected payoff for three hypothetical sellers with different costs as a function of the seller's bid.
    \item[Utility Heatmap] depicts the expected payoff on an eleven-by-eleven grid of hypothetical seller costs and bids.
\end{description} 

\noindent We hypothesize that visualizations presenting bid-relevant information in payoff space (Utility Curves and Utility Heatmap) will improve bid optimality for boundedly rational agents by reducing the cognitive effort needed to identify the best response. We also examine how two types of feedback influence bidding: (1) feedback on the bidder’s payoff, and (2) feedback on inferred costs assuming utility-maximizing behavior. We expect that inferred cost feedback will further improve bidding by helping participants recognize when they have over- or underbid.

We observe statistically significant variation in bid optimization across visualization types despite persistent undershading. Participants using the Allocation Rule visualization achieved the lowest bid optimization accuracy~(ratio between the participant's utility given their bid and the optimal utility), while the Utility Heatmap performed best.  Utility Curves yielded slightly lower bid optimization accuracy than Heatmaps, likely due to fewer reference points in payoff space, though the difference is not significant. Even when most of the computational complexities in determining optimal bidding strategies are removed, participants consistently undershade,\footnote{In the reverse auction we study they underbid this is consistent with prior findings in forward auctions showing persistent overbidding.}.

Further analysis reveals that many participants submitted bids that were effectively insensitive to the competitive environment.
Across experimental conditions, their bids on average clustered near their private cost, with little adjustment in response to differences in the implied win probability or expected payoff. This pattern suggests that participants relied on simple heuristics rather than engaging with the information in the dashboard.  Participants’ self-reported strategies further support this interpretation; several participants reported an implicit preference for wins, which were guaranteed by bidding closer to their costs. 

To better understand whether fixed bidding reflects difficulty interpreting the dashboards or a deeper behavioral tendency, we conducted a follow-up study that introduced training in a deterministic environment where outcomes could be verified.
If undershading was the result of limited visualization literacy, then training effects should carry over to the setting with stochastic payoffs.
If, however, undershading reflects a preference for win certainty, then gains from training should diminish once uncertainty is reintroduced.
We find support for the latter: training improves bid optimization when payoffs are deterministic, but performance reverts to baseline when outcomes are stochastic.

These behavioral findings motivate our contributions to the accurate inference of agent preferences within dashboard mechanisms. We show that the accuracy of inferring bidders' preferences is enhanced by employing a behavioral model that accommodates imprecise optimization and undershading.  Specifically, the quantal response model, which relaxes the perfect rationality assumption, typically addresses the former, while the risk aversion model, which accounts for bidder preference for more certain outcomes, often accounts for the latter, reducing mean squared error in estimated costs by an order of magnitude. We do not have a stance on whether these models are behaviorally correct; instead, we focus on the improvements to inference that they enable.

\section{Related Work}
\subsection{Behavioral Studies of Auctions}

Research in auction design has traditionally focused on identifying mechanisms that achieve desirable outcomes when bidders employ optimal strategies.
However, there is limited research on how to assist decision-makers in the real world in determining effective bidding strategies. 
Behavioral studies of first-price auctions show that humans experience difficulty choosing actions that optimize their utility and consistently overbid~\citep{cox1983test, dur2018identifying, rees2017mistaken,jalaly2017learning,hassidim2016strategic}. To simplify strategic reasoning under uncertainty, dashboard mechanisms propose visualizing the bid allocation rule to improve bidding behavior in auctions that require complex reasoning under uncertainty. If agents respond rationally, dashboard mechanisms
offer provable allocation and welfare guarantees. This motivates our investigation into how providing bid-relevant visualizations influences bidders' behavior and inference accuracy in strategic mechanisms. While prior work has shown that boundedly rational agents continue to struggle with bid optimization when presented with probabilistic information ~\citep{jalaly2017learning}, whether improved dashboard designs can elicit more optimal bidding remains an open question.

\subsection{Visualizations for Decisions and Inference}
Visualization research on decision-making under uncertainty has explored ways of visualizing distributional information that make variation and uncertainty more salient to users, e.g., through the use of frequency representations like hypothetical outcome plots~\citep{hullman2015hypothetical,zhang2021visualizing} or quantile dot plots \citep{fernandes2018uncertainty,kay2016ish}. However, ~\citet{hullman2018pursuit} argue that even when individuals accurately perceive visualizations of probabilistic prediction, like those depicting the allocation rule, they do not necessarily make optimal decisions. While many evaluations have tended to focus on lower-level perceptual tasks, i.e., probability perception, rather than decision quality in an incentivized task~\citep{hullman2018pursuit}, more recent work has studied the role of visualization in aiding decision-making using incentivized decision-making~\citep[e.g,][]{fernandes2018uncertainty,kale2020visual,bancilhon2020let} and strategic-making decision tasks~\citep[e.g,][]{zhang2024designing,kayongo2021visualization} tasks. Our work goes a step further, studying a case where visualization-aided decisions are used for downstream inference in a mechanism. 
\section{Preliminaries}

We study a single-item reverse auction. In each auction round, the participant endowed with a private cost and a computer opponent simultaneously bid to sell a single item. The participant submits a bid from a set of feasible bids ($b \in B$), while the opponent submits a bid randomly drawn from a normal distribution $x \sim \mathcal{N}(\mu, \sigma^2)$. The seller with the lowest bid wins and is paid their bid. By fixing the computerized agents' bid distribution, we isolate potential errors in bidding to the bid optimization of the agent. This allows us to exclude other sources of bidding error, such as errors in estimating the strategies of other agents, which have been extensively studied in the literature on strategic bidding in auctions (see~\citet{crawford2013structural} for a summary).

\subsection{ Best Response Inference}
The participant best responds by submitting a bid that maximizes their utility given by: 

\begin{equation}
\label{eq:utilmax}
       \argmax_{b\in \mathbf{B} }= \pi(b)=(b-c)\cdot Pr(win|bid=b) 
\end{equation}

\noindent \noindent where $c$ is the sellers participants cost and $b$ is their bid. To infer a participant’s cost, we perform a computational grid search that inverts their submitted bid. We construct the computation grid by discretizing the bid and cost space at a resolution of 0.01 and compute the best response bid corresponding to each cost. The inferred cost is then identified as the value whose best response most closely matches the participant’s submitted bid.

\subsection{Quantal Response Inference}

To improve cost inference, we use the logit-based quantal response model, where bidders choose bids probabilistically based on expected utility. Given a seller with cost $c$ and allocation rule $a$, the probability of choosing bid $b$ is:

\begin{equation}
\label{eq:quantalresponse}
    Pr(b|s)=\frac{e^{\lambda\pi(b|a)}}{\sum_{\alpha \in B}e^{\lambda\pi(\alpha|a)}}
\end{equation}

\noindent where $B$ is the set of feasible bids, $\pi(b|a)$ is the expected utility, and $\lambda$ is a precision parameter. As $\lambda \to \infty$, bids converge to best responses; as $\lambda \to 0$, bids are random. We jointly estimate $\lambda$ and cost via maximum likelihood estimation.

\subsubsection{Quantal Response with Constant Relative Risk Aversion}
We extend the quantal response model to incorporate risk aversion using a constant relative risk aversion (CRRA) utility function:

\begin{equation}
\label{eq:utilityrisk}
     \pi_{\text{ CRRA}}=\frac{(b-c)^{1-r}}{1-r} \cdot Pr(win|bid=b)
\end{equation}

 \noindent where r is the Arrow Pratt coefficient typically estimated to be between 0.3 to 0.7 across numerous auction experiments~\citep{goeree2002quantal,chen1998nonlinear,cox1985experimental,harrison1990risk}. To avoid overfitting, we estimate a single $r$ across all participants rather than allowing individual variation \footnote{Estimating individual $r$ values yields only marginal improvements, likely due to widespread underbidding and limited data per participant.}.
 We estimate the quantal precision $\hat{\lambda}$ and cost $\hat{c}$ via maximum likelihood~\citep{mckelvey1995quantal}, minimizing the squared error between observed and predicted bid probabilities estimated by sampling from a model fitted to agent bids.\footnote{Parameters are estimated using the L-BFGS-B algorithm in R.}

\section{Experiment 1: Baseline Dashboard Evaluation}

\subsection{Design}
We conduct a mixed-design repeated-measures experimental auction to assess how dashboard visualization and feedback design influence bid optimization and inference. Participants, each endowed with the same private cost (85 AC), bid against a computer opponent across multiple rounds. In each round, they view a bid-relevant visualization, submit a bid and receive feedback. We vary three visualization types (Fig.~\ref{fig:teaser}) between subjects, two feedback conditions within subjects, and ten allocation rules within subjects.

\subsubsection{Visualization}
We vary three visualization conditions between subjects: Allocation Rules, Utility Curves, and Utility Heatmaps (Fig.~\ref{fig:teaser}). 

\vspace{3pt}
\noindent\textbf{Allocation Rule} visualizes the $Pr(win|bid=b)$ based on the computer opponent’s Cumulative Distribution Function~($F_B(b)$). In each round of our experimental auction, participants bid against a single computer opponent who draws bids from a normal distribution. Given a participant bid $b$, the probability that they win is given by $F_B(b)=Pr(B \ge b)$. To bid optimize, given the allocation rule $ Pr(win|bid=b)$, participants need to calculate their utility for each bid~$b$, given the endowed cost~$c$, and select the bid that maximizes their utility~$\pi$ in expectation. Best response inference assumes that agents accurately compute their expected utility, given the visualized allocation rule. However, analytically calculating the best response bid using probabilities is a challenging task for behavioral agents.

\vspace{5pt}

\noindent\textbf{Utility Curves} visualize the expected utility for hypothetical sellers with different costs$(\{50,75,100\})$ at given bids~(Fig.~\ref{fig:teaser}).  
Participants endowed with a cost of 85 AC can use the visualization to approximate an optimal bid between the utility-maximizing bids for hypothetical sellers with costs of 75 AC and 100 AC.

\vspace{5pt}

\noindent\textbf{Utility Heatmap} visualizes the expected utility for hypothetical sellers with different costs at given bids using color encoding in a data matrix.  To generate the heatmap, we discretize the continuous bid and cost space into fixed intervals $\{50,60,70, \cdots, 150\}$. The smaller interval size in the Utility Heatmap visualizes the expected utility at a higher level of granularity than the Utility Curves while not directly depicting the optimal bid for the participant's endowed cost. Ideally, given the heatmap, participants with an endowed cost of 85 AC submit a bid between the utility optimal bids of sellers with costs of 80 AC and 90 AC~(Fig.~\ref{fig:teaser}).

\vspace{0.5em}

\noindent We hypothesize that, relative to CDFs, the Utility Curves and Utility Heatmap visualizations will improve bid optimization by presenting bid-relevant information in utility space, reducing the cognitive demands of bid optimization to a simpler approximation task. Additionally,  we hypothesize that relative to the Utility Curves, the Utility Heatmap will enable participants to approximate the best response bid more accurately since it provides closer reference points to their endowed cost.

\subsubsection{Feedback}

Feedback has been shown to have a significant behavioral effect on bidding in First Price Auctions (FPAs), especially when it clarifies foregone payoffs~\citep{ockenfels2005impulse, neugebauer2006individual}.
Prior work also shows that presenting probabilities alone does not improve bidding without informative feedback~\citep{ratan2015does}.
Based on these findings, we test two feedback conditions within subjects. In one condition, participants receive feedback on whether they won and their payoff if they win. In the second condition, they also receive feedback on their inferred cost, assuming a best response.
The role of the second condition is to determine whether participants' ability to best respond improves when they receive feedback on the optimality of their bids. An inferred cost higher (resp. lower) than the endowed cost implies overbidding (resp. underbidding).

\subsubsection{ Stimuli Generation}
We aim to meet two criteria when choosing the competing sellers' bid distributions: (1) they generate \textit{varied best response} strategies in different auction rounds, and (2) for any competing bid distributions, the allocation rule is monotone to enable the inversion of bids to obtain costs during inference. To achieve this, we generate the competing agents' bid distribution using a normal distribution and vary the parameters each round. The normal distribution's smooth, monotonically increasing CDF ensures that the visualized allocation is invertible.

\begin{figure*}[h]
 \centering
  \includegraphics[width=\linewidth]{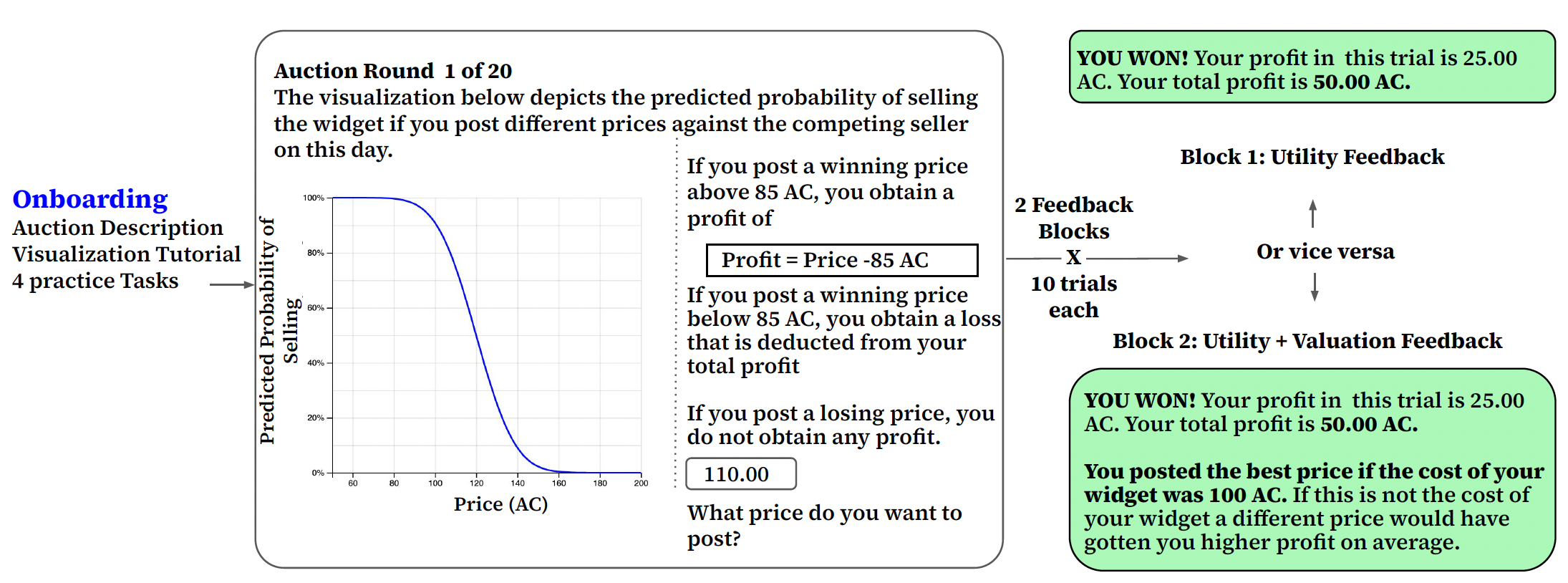}
   \caption{Experimental Auction Study Overview and Interface}
   \label{fig:studyinterface}
 \end{figure*}

\subsubsection{Experimental Procedure}

Participants were randomly assigned to a visualization condition and introduced to the auction scenario and bonus structure \footnote{The bonus was selected based on~\citet{wu2023rational} that proposes that the incentive to consult a visualization in experimental decision-making tasks should be quantified as the difference between the optimal utility obtained given access to visualization $\pi(b_{opt})$ and the baseline utility obtained by a Bayesian rational agent who only has prior information on the distribution of the allocation rules $\pi(b_{prior})$. We incentivize agents to use the visualization by setting payments such that the rational optimal behavior pays 35\% more than the baseline behavior.}. They then completed a brief tutorial on how to interpret their assigned visualization, followed by four training tasks —either making probability judgments (Allocation Rule) or estimating utility (Curves/Heatmap). 
After training, participants completed two blocks of 10 randomized auction rounds. 
In one block, participants receive feedback on the outcome and payoff; in the other block, they are also given feedback on their inferred costs, according to best response. 
The block order was randomized across participants to control for potential order effects. 
Each auction round contains a bidding dashboard with a visualization of bid-relevant information. 
The data visualized to participants is framed as a prediction of the expected outcome given past observations of the bidding behavior of the computer opponent. 
Participants can submit bids in auction coin (AC) units with no minimum \footnote{Although bids below cost were allowed to measure task misunderstanding, we exclude bids yielding negative utility from the analysis. Replicating findings from previous first-price auctions studies ~\citep{kagel1987information}  bid resulting in negative utility occur in roughly 10\% of trials.}. On the last trial in each block, we ask participants to provide a rationale for their submitted bid.

\subsection{Analysis}

We quantify the effect of dashboard design on bid optimality using the bid optimization ratio: 

$$\textit{Bid Optimization Ratio}=\frac{\pi_{bid}}{\pi_{OPT}}$$

\noindent where $\pi_{bid}$ is the participant’s expected utility given their bid, and $\pi_{OPT}$ is the expected utility of the optimal (best response) bid for their endowed cost.

We fit the following preregistered\footnote{Prergistration \url{https://aspredicted.org/8WZ_JBT}} Bayesian hierarchical model to estimate the effects of experimental manipulations on bid accuracy, quantified through the bid optimization ratio:

\begin{equation} \label{eq:bidopt}
\begin{split}
\textit{Bid Optimization Ratio} & = vis*feedback \text{ + } \textit{best response} \text{ + } \\ \textit{block order} \text{ + }
 & trialnum \text{ + } (1 \text{ + }trialnum|participant)  
\end{split}
\end{equation}

\noindent where \textit{vis} is the visualization condition,  \textit{feedback} is an indicator variable indicating the feedback type, \textit{best response} is the optimal bid given the allocation rule in the auction round, \textit{block order} is an indicator variable for the order of the feedback blocks, \textit{trialnum} is an index of the trial number which accounts for learning effects, and \textit{participant} is a unique participant id for modeling individual random effects. We
include an interaction term \textit{(vis * feedback)} to
measure if the bid optimization ratio varies as a joint function of the
visualization and the feedback condition.
We use a zero-one-inflated beta (ZOIB) prior~\citep{brms2017} as the bid optimization ratio is bounded between 0 and 1 and frequently takes on exact values at both extremes.
Effects are evaluated via posterior credible intervals (CIs); non-overlapping CIs between conditions indicate reliable differences.
We aimed to recruit $\sim700$ participants to ensure 80\% power to detect a 3\% difference in bid optimization between visualization conditions.

\subsection{Results}

\begin{figure*}[h!]
 \centering
  \includegraphics[width=\linewidth]{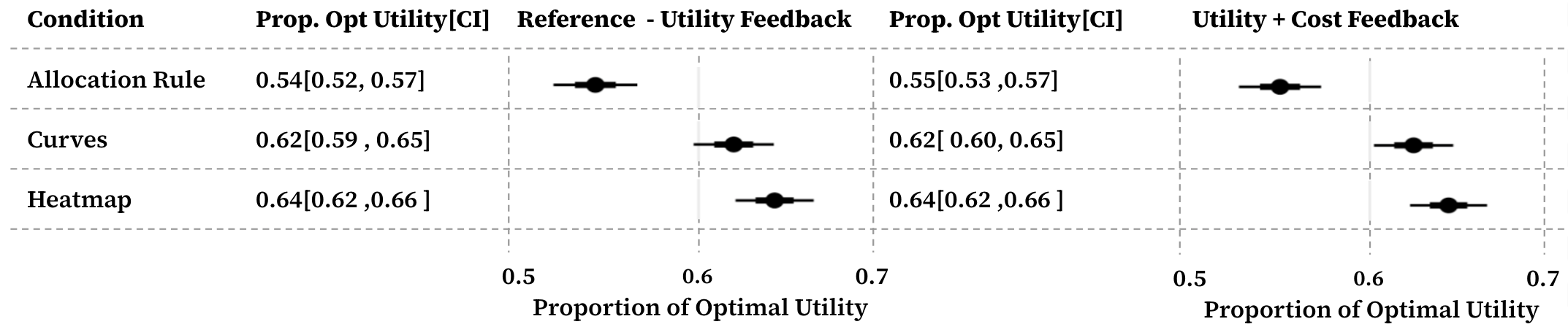}
   \caption{ Bayesian Hierarchical model estimates of the proportion of optimal utility achieved across three visualization conditions—Allocation Rule, Utility Curves, and Utility Heatmap—under two feedback conditions: Utility Feedback (left panel) and Utility + Cost Feedback (right panel). Each dot represents the estimated mean proportion of optimal utility, and horizontal lines indicate 95\% credible intervals. Curves and Heatmap conditions lead to significantly higher bid optimization than Allocation Rule across both feedback types, with minimal additional improvement from cost feedback.}
   \label{fig:bipoptresults}
 \end{figure*}

\subsubsection{Data Preliminaries}

We recruited 704 participants from Prolific\footnote {\url{https://www.prolific.co}}. The median completion time was approximately 13 minutes, with an average base payment of \$18.77 per hour (excluding bonuses).

\subsubsection{Bid Optimization}

We report the conditional means derived from the model estimates in Figure~\ref{fig:bipoptresults}. These conditional means communicate the average bid optimization ratio for visualizations and feedback conditions. Relative to the baseline Allocation rule visualization, the Utility Heatmap and Utility Curves led to significantly higher average bid optimization ratios. 
However, the 95\% Confidence Intervals for the Utility Curves and the  Utility Heatmap visualization conditions overlap, indicating that the improvement to bid optimization of the Heatmap over Utility Curves is not statistically significant. 
Although we hypothesized that providing feedback on inferred costs would improve bidding, it did not result in a statistically significant increase in the bid optimization ratio when compared to the baseline utility feedback condition. 
The absence of an inferred cost feedback effect suggests that the information it provides was dominated by the dashboard’s provision of this information. 
Despite our initial hypothesis that this feedback condition might address overbidding, typical in first-price auctions, we find that it does not. Additionally, even with the best performing dashboard, participants only obtain $\sim$ 64\% of the optimal utility suggesting that  dashboard design choices so not fully eliminate suboptimal behavior, which may reflect not only cognitive limitations but also bidder preferences.
Participants’ self-reported strategies suggest that some undershade--submit lower than optimal bids-- intentionally to increase their chances of winning. Notably, a subset of participants using the Utility Curves and Utility Heatmap visualizations reported recognizing the optimal bid but choosing to submit lower bids anyway, indicating an awareness of their own deviation from optimal behavior.

 \subsubsection{Post Hoc Analysis: Bayesian Rational Agent Analysis}

To evaluate participant performance, we apply the {\em Bayesian rational agent framework} introduced by~\citet{wu2023rational}, which defines a set of normative and diagnostic reference points for decision-making with visualization under uncertainty. The framework compares observed behavior against a family of hypothetical agents who differ in what information they access and how optimally they respond. These comparisons help disentangle whether suboptimal behavior arises from poor information use or imprecise decision-making.

\begin{figure*}[h!]
 \centering
  \includegraphics[width=\linewidth]{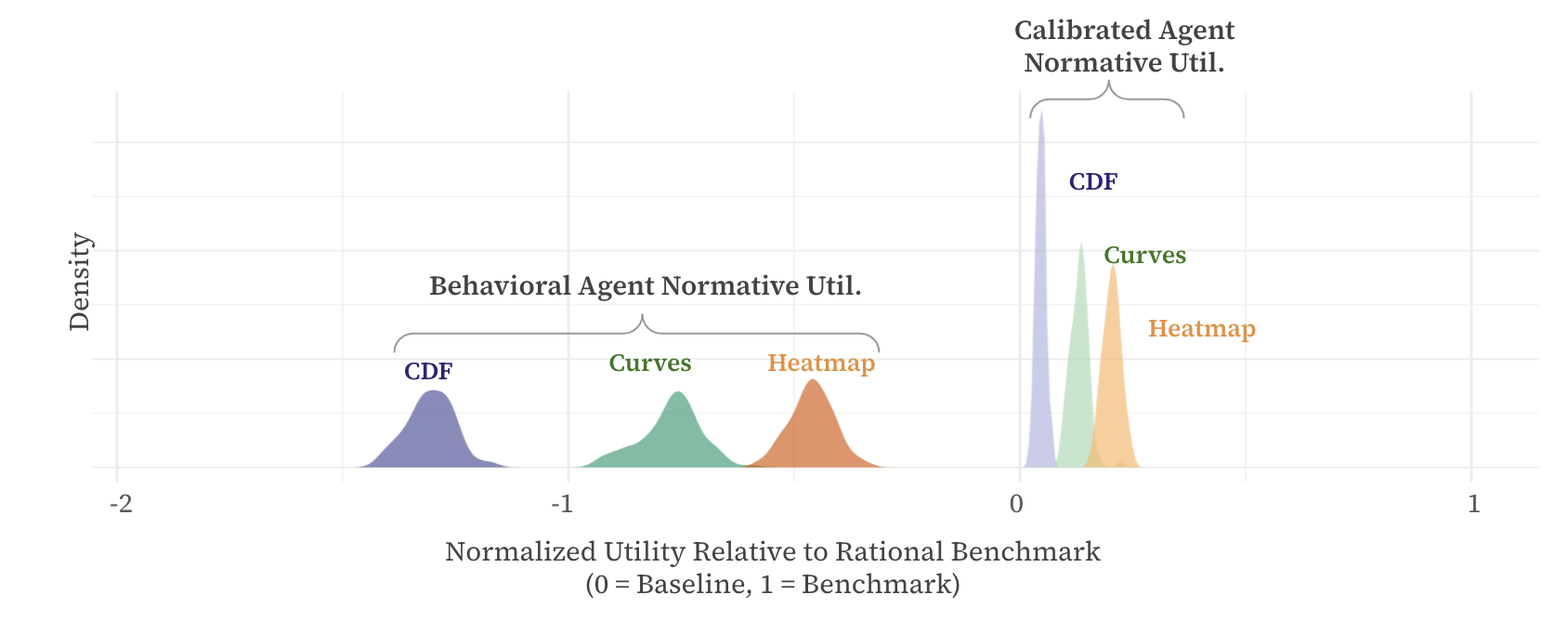}
   \caption{ Normalized utility scores for behavioral and calibrated agents across visualization conditions. Scores are normalized such that 0 represents the baseline agent (prior only) and 1 represents the benchmark agent }
   \label{fig:bayesiananalysis}
 \end{figure*}

Specifically, we compare participants' behavioral outcomes against three hypothetical rational agents: the {\em baseline rational agent},  the {\em benchmark rational agent}, and the {\em calibrated rational agent}. The baseline agent optimizes bids using only the prior distribution over competitor bids. This agent serves as a normative lower bound on rational, utility-maximizing behavior without dashboards. The benchmark rational agent represents an idealized upper bound: a Bayesian agent who observes the dashboard visualization, correctly interprets its informational content, and selects the bid that maximizes expected utility given the visualized distribution over competitor bids. The calibrated rational agent optimally responds not to the true state but to the belief distribution implicitly revealed by participants’ observed bids. In other words, we treat each participant's bid as if it reflects a belief about the underlying competitive environment (e.g., win probability), and we ask: ``What payoff would a rational agent achieve if they held that belief and bid optimally?''. The calibrated agents' payoffs always lie between the performance of the baseline agent and the benchmark agent. The calibrated agent allows us to assess whether participants are meaningfully using the information provided by the dashboard. If behavioral performance is poor but calibrated performance is high, it suggests participants extracted useful information but failed to optimize. If both are low, it indicates a failure to incorporate or interpret the information altogether. In~\citet{wu2023rational}, the performance of the benchmark rational agent is normalized to 1, the performance of the baseline agent is normalized to 0, and the performance of all other agents (behavioral, calibrated) is expressed relative to that benchmark.

Figure~\ref{fig:bayesiananalysis} depicts the normalized score for behavioral and calibrated agents across visualization conditions. We find that participants underperform relative to the baseline Bayesian rational agent (score normalized to 0) in all visualization conditions. Moreover, since participants bid in a manner that is insensitive to the signal, the scores achieved by the calibrated agents---who optimally act on beliefs implicit in participant bids---only marginally exceed those of the baseline agent. This suggests that participants often fail to extract or utilize meaningful information from the dashboard. Together, these results raise doubts about the effectiveness of dashboards alone in improving strategic behavior.

\subsubsection{Inference}

\begin{figure}[t] 
 \centering
  \includegraphics[width=\linewidth]{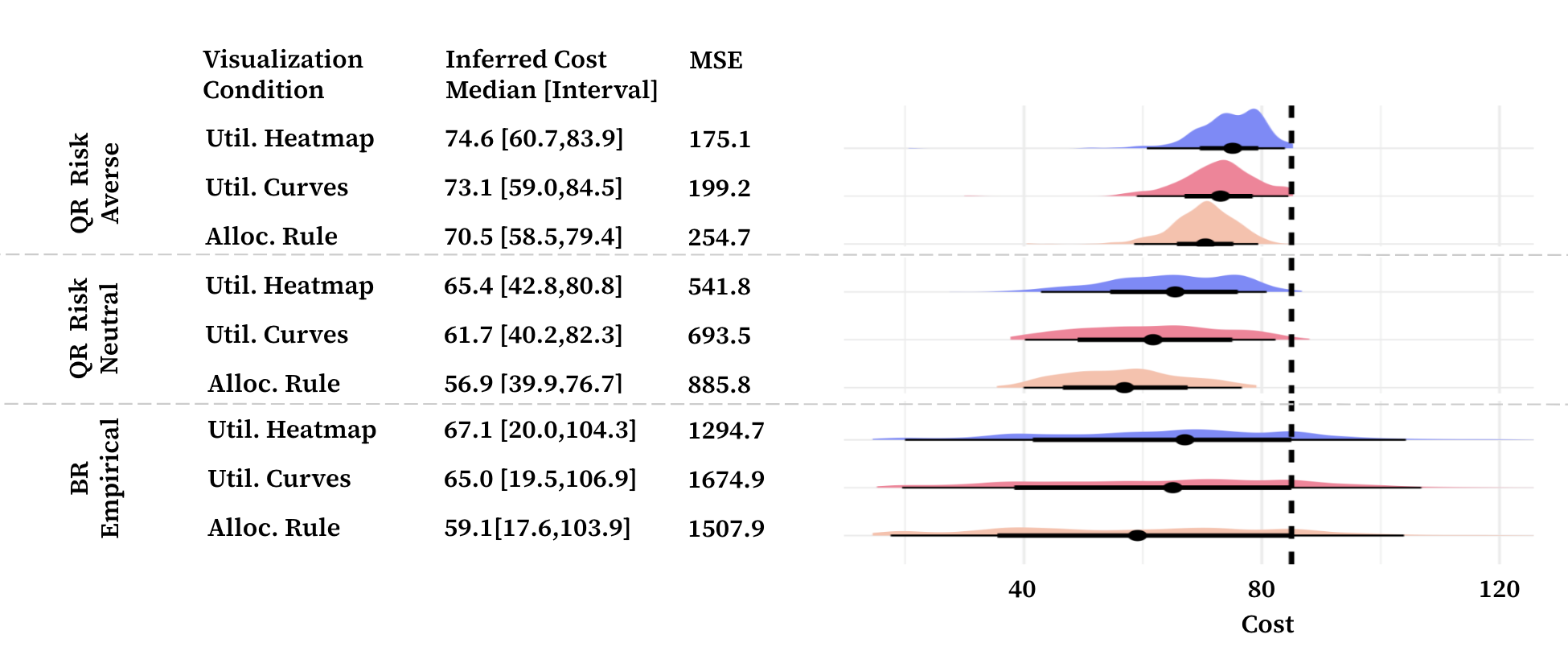}
   \caption{ Distribution of inferred costs using different models and behavioral assumptions ~(BR-Best Response and QR-Quantal Response). The vertical dotted line represents the endowed cost~(85~AC).} 
   \label{fig:gridcostinferencebestresponse}
 \end{figure}

Figure~\ref{fig:gridcostinferencebestresponse} depicts the distribution of inferred costs for different visualizations and models and the corresponding Mean Square Error (MSE) point estimates. 
Under the best response assumption, inferred costs systematically underestimate the true endowed cost (85~AC) and exhibit high variance, resulting in the highest MSE values.
The risk-neutral quantal response model, which relaxes the assumption of perfectly optimal bidding, reduces estimation variance and achieves lower MSE relative to the best response model. However, it still underestimates true costs, as it does not fully account for systematic undershading, which is observed in approximately 80\% of trials.
Incorporating risk aversion into the quantal response model yields substantially more accurate cost estimates with lower variance, reducing MSE by an order of magnitude relative to best response inference.
Nonetheless, even the risk-averse quantal response model continues to underestimate costs, albeit to a lesser extent than both the best response and risk-neutral quantal response models.
Across all models, visualizations that supported more accurate bidding, namely, the Utility Curves and Utility Heatmap, consistently produced the lowest inference error, underscoring the complementary roles of effective visualization design and behaviorally informed inference models.

\section{Experiment 2:  Deterministic Training and Stochastic Transfer}
To isolate the sources of suboptimal bidding observed in Experiment 1, we run a follow-up experiment that tests whether training with verifiable dashboards in a deterministic environment improves bidding under uncertainty. To allow comparison with Experiment 1, participants again complete 20 reverse auction trials (two blocks of 10) using the same allocation rules and fixed cost (85 AC). However, three key changes are introduced: (1) feedback is held constant—all participants receive outcome and payoff feedback; (2) the first block features training trials with deterministic payoffs, allowing participants to verify the dashboard's accuracy, while the second block reintroduces stochastic payoffs to assess transfer; and (3) visualization type is manipulated only during training. Participants were randomized to one of three training conditions:

\begin{description}
\item \textbf{Deterministic True-Cost Utility Curve}: visualized a utility curve for the participant’s cost (85 AC) in all 10 training trials.
\item \textbf{Deterministic Utility Curves}: visualized utility curves for hypothetical sellers with different costs as a function of the seller’s bid~(as in Experiment 1) in all 10 training trials.
\item \textbf{Deterministic Combined True-Cost Curve and Hypothetical Curves}: visualized the True-Cost Curve for the first 5 trials and the Hypothetical Seller Utility Curves for the remaining 5.
\end{description}

\noindent In all trials with stochastic payoffs, participants view utility curves for hypothetical bidders, which improved bidding in Experiment 1 and generalizes naturally from the curve for the participants' true cost used in the training.

Beyond testing whether training with dashboards in deterministic settings improves bidding under uncertainty, Experiment 2 isolated additional sources of suboptimality. Persistent undershading in the deterministic setting after exposure to the True-Cost Utility Curve may indicate failures in learning or attention, while undershading with Approximate Utility Curves in deterministic settings suggests difficulty identifying an optimal bid from inexact stimuli. 
Finally, differences in performance between deterministic and stochastic blocks may indicate the impact of uncertainty on bidder behavior.

We fit a Bayesian hierarchical model to estimate the effects of training on bid accuracy, measured by the bid optimization ratio. The model includes fixed effects for the interaction between training phase and visualization type, with a random intercept for participants. The model is specified as follows:

\begin{equation} \label{eq:bidoptstudy2}
\begin{split}
\textit{Bid Optimization Ratio} & \sim training*vis  + (1 |participant)  
\end{split}
\end{equation}

\noindent where \textit{training} is the training condition,  \textit{vis} is the visualization type and \textit{participant} is a unique participant id for modeling individual random effects. 

We report model-based conditional means of the bid optimization ratio by training condition and visualization type. These estimates reflect the average proportion of the optimal utility that participants obtained for a given condition.

\subsection{Results}

\subsubsection{Data Preliminaries}

We recruited 240 participants from Prolific. The median completion time was approximately 17 minutes, with an average base payment of \$14.02 per hour (excluding bonuses).

\subsubsection{Bid Optimization}
\begin{figure*}[h!]
 \centering
  \includegraphics[width=\linewidth]{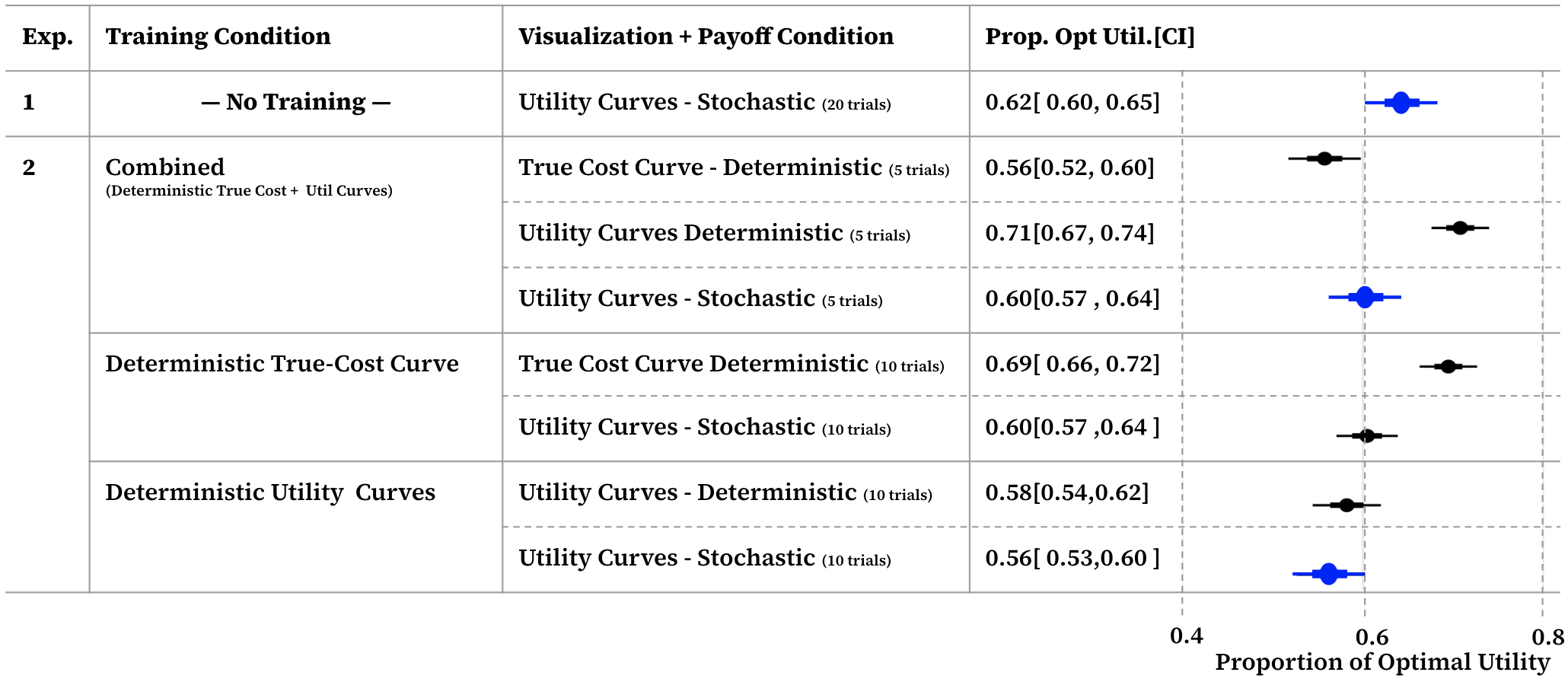}
   \caption{ Bayesian hierarchical model posterior estimates of the proportion of optimal utility (bid optimization ratio) achieved with Utility Curve visualizations in Experiment 1 and across all training and payoff conditions in Experiment 2. Results reflect mean performance with 95\% credibility intervals. }
   \label{fig:study2results}
 \end{figure*}

As shown in Fig.~\ref{fig:study2results}, participants who trained exclusively on their true-cost curve, as well as those in the combined condition (training first on true-cost, then on approximate utility curves), demonstrated improved bid optimization during training, achieving approximately 71\% of the utility-optimal welfare. However, across all training conditions, participants’ performance when using Utility Curves in the stochastic test trials remained similar to that observed in Experiment 1.) This lack of performance improvements may reflect the persistence of individual-level risk preferences or limits in generalizing from deterministic training to stochastic test conditions. Additionally, participants’ inability to maximize payoffs even when shown cost-specific curves in deterministic trials suggests possible inattention or inaccurate mental models of auction dynamics.
Together with the results from the first experiment , these findings suggest that visualization dashboards, while helpful in most environments, may be insufficient for promoting strategic behavior.

\section{Discussion and Future Work}

Our research findings carry implications for strategic environments that employ visualization dashboards to aid agent decision-making.
While dashboards can improve decision-making, especially for boundedly rational agents, they do not eliminate suboptimal behavior.
In our reverse auction study, visualizations such as Utility Curves and Utility Heatmaps significantly improved bid optimization compared to normative Allocation Rule dashboards.
Yet even with improved visualizations dashboards and training on how to use them in verifiable contexts, bounded rationality and a preference for wins led to persistent undershading. A key performance benchmark in our study is the bid behavior of a Bayesian rational agent with access to the prior distribution over competitor bids but no dashboard~\citet{wu2023rational}. This agent represents a lower bound on rational behavior in the absence of information offered by the dashboard. Surprisingly, despite having access to visualized payoff information, participants consistently earned less utility on average than the Bayesian rational agent with only the prior. This underperformance persisted even with improved dashboard design, training, and deterministic payoffs.
These findings reveal a practical limitation of dashboard mechanisms~\citep{hartline2019dashboard, deng2021welfare}: merely publishing a transparent dashboard is insufficient to ensure near-optimal agent behavior or accurate inference. 
In particular, assuming agents respond optimally (i.e., best response) can lead to significant inference errors. 
Incorporating behavioral models, such as quantal response with risk aversion, yields more accurate inference and should be the default approach when designing dashboard mechanisms.

An important direction for future work is to evaluate whether these findings generalize to strategically sophisticated agents or higher-stakes environments. Our study focused on novice participants and modest incentives, providing insight into how boundedly rational agents interact with dashboard mechanisms.
For example, Allocation Rule dashboards may be more effective for expert bidders with greater strategic sophistication.
However, conducting large-scale, well-powered experiments with expert participants presents practical challenges and remains an open area for future exploration.
Additionally, while our study focused on dashboards that truthfully implement the published allocation rule, in real-world platforms, such consistency may not always hold due to auction dynamics.
An important direction for future work is to empirically evaluate how agents respond to imperfect dashboards.
Prior theoretical work~\citep{hartline2019dashboard, deng2021welfare} proposes methods to mitigate such deviations, but empirical validation remains an open question.
Future work could explore alternative decision models (e.g., prospect theory, loss aversion), adaptive dashboards that personalize to user behavior, and new experimental paradigms to test dashboard mechanisms under more realistic, dynamic market conditions. Additionally, given the weak responsiveness to varied visualized signals observed in our study, models that assume fixed bid shading or weakly signal-dependent shading—such as constant or linear shading in signal space—may better capture observed deviations from optimal bidding and improve the accuracy of value inference in behavioral settings.

\section{Limitations \& Future Work}

The generalizability of our results may be limited by the choice of experimental design (e.g., participant selection and reward amounts). Our study was conducted on novice participants who may have lacked experience with auctions. To address this problem, we performed a preregistered exclusion of participants who consistently bid lower than their cost, as it indicated that they misunderstood the basic auction premise. However, other sources of misunderstanding, i.e., misunderstanding the scoring rule, may have persisted. It would be interesting to see if there were significant differences in effects with savvy bidders, i.e., would a visualization like the Allocation Rule, which obfuscates less information, work better for them? However, conducting a  lab study with savvy bidders at sample sizes with sufficient statistical power may be challenging. The small monetary bonuses obtained by participants in the task may not have served to incentivize them sufficiently; thus, the results observed in our study may differ from those observed in practice when agent monetary costs and rewards are larger.

Effective dashboard mechanism design aims to incentivize agents to best respond by implementing the allocations they publish. Our study evaluated bidding in this case, where best response is the utility-optimal response for each agent. However, in practice, publishing such a dashboard may be challenging due to unpredictable auction dynamics. Theoretical work on dashboard mechanisms proposes solutions to address sub-optimal auction outcomes caused by deviations between the published dashboards and the allocations obtained after running the mechanism. ~\citet{hartline2019dashboard} track over and undercharges due to unpredictable dynamics and adjust the dashboard in future rounds accordingly. \citet{deng2021welfare} assume stable agent value profiles between rounds in repeated auctions, e.g., via a bounded vector norm, and use a conservative dashboard to ensure allocation feasibility. However, the empirical evaluation of how agents respond to dashboards that do not implement the rules they publish in each auction round remains an open problem.

Accurate cost prediction is fundamental to designing practically successful dashboard mechanisms. Our research indicates that value inference is not highly accurate, necessitating further work to understand how to create effective dashboards and more accurate inference methods for the practical success of the approach proposed in~\citet{hartline2019dashboard} and ~\citet{deng2021welfare}. In light of bidder behavior in our study, future work may explore alternative decision-making models under uncertainty, i.e., prospect theory, loss aversion, and pessimistic decision-making, as potential avenues for improving inference accuracy.

\bibliography{biblio}

\appendix

\end{document}